\documentclass[preprint]{aastex}

\shorttitle{Modes of star formation}
\shortauthors{Adams \& Myers}

\newcommand{\be}{\begin{equation}}
\newcommand{\ee}{\end{equation}}

\newcommand{\mcone}{{\langle M_C \rangle}} 
 
\newcommand{\fun}{ {\cal F}_{\ast}}
\newcommand{\bmin}{b_{\rm min} }
\newcommand{\tdk}{ t_{\rm disk}} 
\newcommand{\ec}{ {\epsilon}_{C}}
\newcommand{\fuv}{ {\langle \langle F_{uv} \rangle \rangle }}

\begin{document}

\title{Modes of Multiple Star Formation} 

\author{Fred C. Adams} 

\affil{Michigan Center for Theoretical Physics} 
\affil{Physics Department, University of Michigan, Ann Arbor, MI 48109} 

\email{fca@umich.edu} 

\author{Philip C. Myers} 

\affil{Harvard-Smithsonian Center for Astrophysics, 60 Garden Street, 
Cambridge, MA 02138} 

\email{pmyers@cfa.harvard.edu} 

\begin{abstract} 

This paper argues that star forming environments should be classified
into finer divisions than the traditional isolated and clustered
modes. Using the observed set of galactic open clusters and
theoretical considerations regarding cluster formation, we estimate
the fraction of star formation that takes place within clusters. We
find that less than $\sim10\%$ of the stellar population originates
from star forming regions destined to become open clusters, confirming
earlier estimates. The smallest clusters included in the observational
surveys (having at least $N\sim100$ members) roughly coincide with the
smallest stellar systems that are expected to evolve as clusters in a
dynamical sense. We show that stellar systems with too few members $N
< N_\star$ have dynamical relaxation times that are shorter than their
formation times ($\sim1-2$ Myr), where the critical number of stars
$N_\star$ $\approx$ 100. Our results suggest that star formation can
be characterized by (at least) three principal modes: [{\bf I}]
isolated singles and binaries, [{\bf II}] groups ($N<N_\star$), and
[{\bf III}] clusters ($N>N_\star$). Many -- if not most -- stars form
through the intermediate mode in stellar groups with $10<N<100$. Such
groups evolve and disperse much more rapidly than do open clusters;
groups also have a low probability of containing massive stars and are
unaffected by supernovae and intense ultraviolet radiation fields.
Because of their short lifetimes and small stellar membership, groups
have relatively little effect on the star formation process (on
average) compared to larger open clusters.

\end{abstract}


\section{INTRODUCTION} 

The past two decades have witnessed the development of a working
paradigm for the formation of isolated single stars (e.g., see various
reviews from Shu, Adams, \& Lizano 1987 to Mannings, Boss, \& Russell
2000 [PPIV]). In practice, however, stars tend to form in groups and
clusters, and many authors have suggested that clusters form the
majority of stars in our galaxy (e.g., Evans 1999; Elmegreen 1985). If
most star formation takes place within sufficiently dense
environments, then the current theory of star formation could require
substantial modification, or perhaps even a new paradigm. A vital
issue that must be addressed is to determine the distributions of
sizes and stellar densities for star forming systems and to estimate
how neighboring stars and the background environment affect the
formation of a given star. We can view star formation environments --
groups and clusters -- in two different ways: [1] We can consider
these stellar systems as astronomical objects in their own right and
study their birth, evolution, and eventual demise. [2] We can consider
the effects of groups and clusters on the star formation process.
These two viewpoints are intrinsically coupled in that the evolution
of these systems determines, in part, their effects on star
formation. In this work, we necessarily consider both of these points
of view.

For stars born within regions of sufficiently high stellar density,
many possible effects can influence their formation and subsequent
development. If stars form in a stellar system that lives for many
dynamical times during the formation stage, then processes such as
mergers can affect stellar formation (e.g., Murray \& Lin 1996;
Bonnell et al. 1997, 1998).  If newly formed stars continue to live
within a dense cluster-like environment, their final characteristics
are influenced by additional effects, including binary capture through
disk dissipation (Heller 1993; Ostriker 1994), scattering of planets
into more eccentric orbits (Laughlin \& Adams 1998), and other related
dynamical events (see also Price \& Podsiadlowski 1995; Allen \&
Bastien 1995). Sufficiently large clusters are likely to produce
massive stars, which can affect star formation through their intense
ultraviolet radiation fields (St{\"o}rzer \& Hollenbach 1999; Hester
et al. 1999) and supernova blast waves (e.g., Cameron \& Truran 1977).

To assess the importance of these processes, we need to identify the
types of stellar systems that form stars and determine the effects of
these environments on star formation. Stars form within stellar
systems containing $N$ members, where $N$ is known to vary from 1 to
$10^4$. The isolated mode of star formation refers to the limit $N=1$
(or 2) and allows no influence of neighboring stars.  The clustered
mode of star formation refers to the limit $N \gg 1$ and allows for
strong influences of neighboring stars. Most stars form in systems
with $N > 1$, but we still seek answers to the following questions:
What is the typical size $\langle N \rangle$ and the distribution of
system sizes $P(N)$? How does a system of size $N$ affect its
constituent stars? Is the typical star forming system more nearly
isolated or clustered?

In this paper, we take a modest step toward these goals. We first
estimate the fraction of star formation that takes place in clusters
versus smaller stellar systems that we denote as groups.  Along the
way, we make a clean distinction between systems with large numbers
$N$ of stars (clusters) and those with smaller $N$ (groups).
Specifically, we estimate the critical number of stars $N_\star
\approx 100$ that defines the boundary between these two types of star
forming systems. Some distinction between groups and clusters has long
been known from observations, especially from infrared imaging of
embedded regions.  For example, in Orion B (L1630), stars are
organized into clusters (Lada et al. 1991) with relatively few groups
or distributed (isolated) young stellar objects (Li, Evans \& Lada
1997); in contrast, the southern part of Orion A (L1641) contains many
groups and distributed stars (Strom et al. 1993, 1994; Chen \&
Tokunaga 1994).

After making a distinction between clusters and smaller groups, we
must assess the effects of these different environments on star
formation.  Stars born within clusters have a chance to experience
disruptive close encounters with other stars, whereas stars born in
smaller groups have much smaller odds of such interactions. In this
context, close encounters are those that lead to substantial
destructive or constructive effects on nascent solar systems (such as
binary capture or planetary ejection); all stellar groups evolve
through scattering encounters that lead to their eventual
dispersal. Because of statistical considerations and the form of the
stellar initial mass function (IMF), clusters are large enough to have
an appreciable chance of containing massive stars; smaller groups, on
average, contain only low mass stars. Stars forming within clusters
are thus influenced by nearby supernova explosions and intense
ultraviolet radiation fields; stars forming within smaller groups are
relatively unaffected by massive stars and their destructive effects.

This paper is organized as follows. Using observed surveys of galactic
open clusters, we estimate (in \S 2) the fraction of stars that are
born within clusters. In \S 3, we find the minimum number $N_\star$ of
stars required for a stellar system to have its dynamical relaxation
time longer than its formation time; only those systems above the
threshold $N > N_\star$ live long enough ($\sim100$ Myr) to be
considered as clusters. In \S 4, we discuss how groups and clusters
affect the star formation process through stellar scattering events,
supernovae, and background radiation fields. We conclude, in \S 5,
with a summary and discussion of our results.

\section{FRACTION OF STAR FORMATION IN CLUSTERS} 

In this section, we estimate the fraction of stars that are initially
born within open clusters. Many authors have studied the distribution
of open clusters in our galaxy and this fraction is relatively well
known ($f_C \sim 0.1$).  In this present discussion, we estimate $f_C$
using the observational survey of Battinelli \& Capuzzo-Dolcetta
(1991; hereafter BC91), who selected a collection of 100 open clusters
in the solar neighborhood [see also the analyses of van den Bergh
(1981); Elmegreen \& Clemens (1985); Pandey \& Mahra 1986; Janes et
al. (1988)]. The principal result of the BC91 survey is the cluster
formation rate ${\cal R}$,
\be 
{\cal R} = 0.45 \pm 0.04 \, \, {\rm Myr}^{-1} \, \, {\rm kpc}^{-2} \, . 
\label{eq:cfrate} 
\ee 
The cluster formation rates from all of the observational surveys are
roughly consistent with each other (at the factor of two level) and
the BC91 result is among the highest.

To proceed further, we specify the time dependence of the star
formation rate, which we take to have an exponentially decreasing
form, $\propto$ e$^{-qt}$, over the age of the galactic disk. The
total number of clusters, per kpc$^2$, produced over the age of 
the disk is thus 
\be
N_C = \int_0^{\tdk} {\cal R} {\rm e}^{q (\tdk - t)} \, dt \, = 
{ {\cal R} \over q} \bigl\{ {\rm e}^{q\tdk} - 1 \bigr\} \, , 
\ee
where $\tdk \sim 9-10$ Gyr (Wood 1992) and where $\cal R$ is the
present day cluster formation rate (given by eq. [\ref{eq:cfrate}]).
The value of $q$ can be estimated from models of the chemical
evolution of the galaxy (Rana 1991) or from the white dwarf luminosity
function (Wood 1992; see also Adams \& Laughlin 1996). The white dwarf
luminosity function can be fit using a constant star formation rate.  
With some uncertainty, the chemical evolution models indicate a slowly
decreasing star formation rate with $q \tdk$ $\approx$ $\ln 4$. Since
we want to determine the largest fraction of star formation that can
take place within clusters, we use the decreasing form with $q \approx
\ln 4/\tdk \approx$ 0.14 Gyr$^{-1}$ as our standard case.

If the average cluster mass is $\mcone$, and if all clusters are
eventually slated for destruction, then clusters contribute a fixed
amount $\Delta \Sigma$ to the surface density of the galactic disk, 
\be
\Delta \Sigma = {\cal R} \, \tdk \,  \mcone \, 
{1 \over q \tdk} \bigl\{ {\rm e}^{q\tdk} - 1 \bigr\} \, . 
\ee
For comparison, the observed surface density of the galactic disk is
26.4 $M_\odot$ pc$^{-2}$ in visible stars, with an additional 18.2
$M_\odot$ pc$^{-2}$ in stellar remnants (Binney \& Tremaine 1987;
hereafter BT87). Correcting for mass loss in the transformation
between progenitor stars and stellar remnants, we obtain a total
stellar surface density of $\Sigma_\ast \approx 63$ $M_\odot$
pc$^{-2}$.  The fraction $f_C$ of the stellar disk component
contributed by open clusters is thus given by 
\be
f_C = { \Delta \Sigma \over \Sigma_\ast} \approx 
7 \times 10^{-5} \, \Bigl( {\mcone \over 1 M_\odot} \Bigr) 
\, {1 \over q \tdk} \bigl\{ {\rm e}^{q\tdk} - 1 \bigr\} \, . 
\label{eq:fcdef} 
\ee
In the BC91 sample, the observed cluster luminosity function implies 
a typical cluster mass of $\mcone$ $\approx$ 500 $M_\odot$ (which is
subject to some uncertainty due to the required transformation from
luminosity to mass). Using this result and $q \tdk$ = $\ln4$, we find
the fraction of stars that form in clusters: $f_C \approx 0.077$.
Thus, open clusters contribute $\sim 8 \%$ of the field stars.
Although this fraction is substantial, open clusters are {\it not} the
birth places for the majority of stars (as sometimes claimed). If we
consider the limiting case of a constant star formation rate, $q \to
0$, equation [\ref{eq:fcdef}] implies a smaller cluster fraction $f_C
\approx$ 0.035.

The above analysis uses results integrated over the age of the
galactic disk. We can obtain a consistency check by using the present
day values. The cluster formation rate [\ref{eq:cfrate}] and $\mcone
\approx 500 M_\odot$ jointly imply a star formation rate in clusters
$(SFR)_C \approx$ 225 $M_\odot$ kpc$^{-2}$ Myr$^{-1}$, whereas the
current star formation rate in the solar neighborhood is substantially
larger, $(SFR)_T$ $\approx$ 3000 -- 5000 $M_\odot$ kpc$^{-2}$
Myr$^{-1}$ (Rana 1991). These present day values thus indicate that
the fraction $f_C$ of star formation that takes place in clusters lies
in the range 0.045 $< f_C <$ 0.075, consistent with the time
integrated estimates found previously.

This result for the fraction of star formation in open clusters 
($f_C < 0.1$) is consistent with previous results.  For example,
Roberts (1957) estimated that 10\% of stars born in the galaxy are
formed within exposed clusters.  Elmegreen \& Clemens (1985) suggest
that 10\% of low mass clouds form bound clusters, with the remaining
clouds producing a distributed stellar population.  Similarly, Lada
(1999) suggests that most embedded clusters emerge from molecular
clouds as unbound systems; although most stars may form in embedded
``clusters'', the majority of these stellar groups evolve to become
unbound associations rather than bound open clusters.

In this discussion, the implied definition of a cluster requires that
the system is big and bright enough to be included in the BC91
analysis. The survey has a limiting absolute V magnitude of --4.5,
which means that only young clusters ($\log_{10} [t/{\rm yr}]$ $\sim
6.5$) more massive than $\sim 500 M_\odot$ are directly included.
However, BC91 use a mass function for the clusters that extends down
to $\sim 100 M_\odot$ to derive their cluster formation rate.  If this
extrapolation were completely accurate, then the cluster formation
rate given by equation [\ref{eq:cfrate}] would include all clusters
larger than $\sim$100 $M_\odot$. Nonetheless, selection effects could
lead to an underestimate of the number of the smallest clusters.  The
value of $f_C$ derived here should thus be considered as the fraction
of ``large clusters'' and may not include all clusters with only a
couple hundred members.  An effective boundary (in $N$) thus separates
the larger clusters included in the observational surveys from smaller
systems which are not included; for the BC91 survey, this boundary is
at $N \sim 300$.

As we show next, sufficiently small stellar systems have dynamical
relaxation times that are shorter than their formation times. As a
result, such small groups of stars are not clusters in a practical
sense. These groups differ from clusters not only in their numbers $N$
of members but also in their physical properties: Clusters evolve
slowly and experience both a collisionless and an interactive phase;
groups evolve quickly and have no collisionless phase. Young groups
and clusters thus represent different types of physical systems (see
\S 3) and have different effects on star formation (see \S 4).

\section{MINIMUM NUMBER OF CLUSTER STARS} 

In this section, we estimate the minimum number $N_\star$ of stars
required for a stellar system to behave as a cluster.  As a general
rule, a system of stars cannot evolve as a cluster unless its 
relaxation time $t_{\rm relax}$ is sufficiently long. For the sake of
definiteness, we require the relaxation time to be longer than the 
cluster formation time $t_{\rm form}$, i.e., 
\be
t_{\rm relax} \ge \alpha \, t_{\rm form} \, , 
\label{eq:alphadef} 
\ee
where $\alpha$ is a dimensionless number greater than unity. We thus
want to find the minimum number $N_\star$ of stars necessary to
satisfy inequality [\ref{eq:alphadef}].

The dynamical relaxation time, which depends on the cluster size $N$,
is the time required for a cluster member to change its velocity by a
relative amount of order unity. For cluster ages shorter than the
relaxation time, $t < t_{\rm relax}$, the stars in a cluster do not
interact appreciably.  For longer times, $t > t_{\rm relax}$, the
effects of interactions add up and the cluster alters its structure.
On longer times, $t \gg t_{\rm relax}$, severe structural changes are
forced upon the cluster due to stellar loss through evaporation,
ejection, and core collapse.  The relaxation time varies over the
structure of the cluster. One usually adopts the median relaxation
time -- that evaluated near the median radius -- as a characteristic
time scale for the system. As a reference point, the typical
evaporation time is of order 100 median relaxation times (BT87). 
The relaxation time can be written in the form  
\be
t_{\rm relax} = Q_{\rm relax} t_{\rm cross} \, , 
\label{eq:trelax} 
\ee
where the crossing time is $t_{\rm cross}$ = $R/v$ and where 
$Q_{\rm relax}$ is the number of crossings required to make the
velocity of a star change by a relative amount comparable to unity.
The velocity $v$ is related to the cluster size $R$ through the 
depth of the cluster potential well, i.e., 
\be
v^2 = {G M \over R} \, , 
\label{eq:vofr} 
\ee
where $M$ is the total mass of cluster, including both stars and
any remaining gas. For a purely stellar system, $Q_{\rm relax}$
$\approx N/10 \ln N$ (BT87).  For the present application, however, 
we generalize this result to include the presence of cluster gas. 
We define $\epsilon$ to be the star formation efficiency of the
cluster, i.e., $\epsilon$ = $N \langle m_\ast \rangle / M$, where
$\langle m_\ast \rangle$ is the mean stellar mass and where $N$ is the
number of cluster stars. The number of crossings per relaxation time
is then given by 
\be 
Q_{\rm relax} \approx 
{N \epsilon^{-2} \over 10 \ln [N/\epsilon]} \, . 
\label{eq:qrelax} 
\ee 

We want to find a lower limit $N_\star$ for the number of stars
required for the system to behave as a cluster.  For a purely stellar
system with $\epsilon = 1$, we see immediately that for sufficiently
small numbers $N$ of stars, $Q_{\rm relax}$ $< 1$, and the relaxation
time is less than the crossing time.  This critical number $N_1$ of
stars (that required for $N / 10 \ln N$ = 1) is $N_1 \approx 36$. A
firm lower bound on the minimum number $N_\star$ of stars required 
for a system to evolve as a cluster is thus given by 
$N_\star > N_1 \approx 36$. 

We can find a more interesting limit by considering the formation of
the cluster and hence by invoking the constraint of equation
[\ref{eq:alphadef}].  Unfortunately, we do not yet have a well
developed theory of cluster formation. To make a start, we assume that
a cluster forms out of the collapse of a molecular cloud. The
formation time must be somewhat longer than the sound crossing time of
the system (Elmegreen 2000), so we write 
\be 
t_{\rm form} \ge \beta \, {r_\infty \over a} \, , 
\label{eq:betadef}
\ee
where $a$ is the effective sound speed, $r_\infty$ is the initial size
of the cloud fragment that forms the cluster, and $\beta$ is a
dimensionless number larger than (but of order) unity. To obtain a
simple model of the initial cloud, we assume that it takes the form of
an isothermal sphere. In this case, the initial cloud size $r_\infty$
is related to the cluster mass $M$ and sound speed through the
relation $r_\infty = GM/2a^2$ (Shu 1977). Notice that any departures
from the isothermal model can be incorporated into the parameter
$\beta$. As an alternate model, e.g., we could specify the cluster
formation time using the free fall collapse time of a uniform density
gaseous sphere. With the proper choice of $\beta$, however, we recover
{\it exactly} the same mathematical form for our derived constraint
[$\beta$(uniform sphere) = $(4/\pi) \gamma^{3/2}$ $\beta$ (isothermal
sphere), where $\gamma$ is defined in equation [\ref{eq:gammadef}]
below].  As a rough estimate, the sound crossing time of the cluster
is $\sim10^6$ yr, an order of magnitude longer than the time scale 
($\sim10^5$ yr) for individual star formation events 
(Myers \& Fuller 1993; Adams \& Fatuzzo 1996). Keep in mind that
$r_\infty$ is the size of the initial mass distribution and hence is
larger than the size $R$ of the newly formed cluster.  We thus define
an additional dimensionless parameter 
\be 
\gamma \equiv {r_\infty \over R} \, , 
\label{eq:gammadef} 
\ee
where we expect $\gamma$ to be larger than (but of order) unity.

Putting all of the above conditions together (eqs. [\ref{eq:alphadef} 
-- \ref{eq:gammadef}]), we derive the constraint 
\be
{N/\epsilon \over \ln [N/\epsilon] } \ge 
10 \, \sqrt{2} \, \alpha \, \beta \, \gamma^{3/2} \, \epsilon \, . 
\label{eq:nlimit} 
\ee 
This constraint (eq. [\ref{eq:nlimit}]) grows weaker as the star
formation efficiency $\epsilon$ decreases; for sufficiently low values
of $\epsilon$, the limit becomes too weak to provide a meaningful
bound.  In order for the cluster to remain gravitationally bound,
however, the star formation efficiency cannot become too small.
Before analyzing this compromise in detail (see below), we obtain a
rough estimate using $\epsilon = 1/2$ as a typical value. We also
adopt $\alpha = 1$ which corresponds to the relaxation time and the
cluster formation time being equal. If we had a definitive theory of
cluster formation, the remaining parameters $\beta$ and $\gamma$ (or
their distributions of allowed values) would be unambiguously
specified. In the absence of a complete theory, we must rely on
estimates and hence we adopt $\beta$ = 2 = $\gamma$. For this case,
the limit becomes $N \ge 108$.  Thus, the minimum number $N_\star$ of
stars required for the cluster formation time to be shorter than the
dynamical relaxation time is $N_\star \sim 100$.

We now derive a more rigorous constraint on $N_\star$ by considering
the whole range of star formation efficiencies. Virial arguments
suggest that clusters remain bound if $\epsilon > 1/2$ and become
unbound if $\epsilon < 1/2$ (for rapid gas removal; see Hills 1980, 
Mathieu 1983, Elmegreen 1983). In practice, however, the stars in a
forming cluster have a distribution of velocities. The low velocity
stars in the tail of the distribution survive as a gravitationally
bound entity even if $\epsilon < 1/2$; the high velocity stars in the
opposite tail escape even if $\epsilon > 1/2$.  A more accurate
description is that a cluster formed with $N$ stars (before gas
removal) eventually produces a bound cluster with $N_f = \fun
(\epsilon) N$ stars after gas removal. The function $\fun (\epsilon)$
varies smoothly with star formation efficiency rather than exhibiting
step function behavior. The shape of $\fun (\epsilon)$ depends on the
shape of the distribution function for the cluster stars, the rate of
gas removal, and the density distributions of the stars and gas (e.g.,
Adams 2000; Kroupa, Aarseth, \& Hurley 2000; Geyer \& Burkert 2000;
Lada, Margulis, \& Dearborn 1984). For clusters with isotropic
velocity distributions, for example, the function $\fun \approx (2
\epsilon - \epsilon^2)$ provides a good approximation over the
expected range of cluster models (see Fig. 3 of Adams 2000).

To incorporate limits on the star formation efficiency into our
analysis, we require that the final bound cluster (which contains
$N_f$ stars after gas removal) have its relaxation time longer than
its crossing time, i.e., $N_f \ge N_1$, where $N_1 \approx 36$ is
defined above.  We thus impose the additional constraint 
\be 
\fun (\epsilon) N \ge N_1 \, .  
\label{eq:elimit} 
\ee 
For a given function $\fun (\epsilon)$, the coupled constraints 
[\ref{eq:nlimit}] and [\ref{eq:elimit}] define a well posed 
optimization problem. 

The solution is straightforward. The first constraint
[\ref{eq:nlimit}] says that $N > f_1 (\epsilon)$ = $\lambda \epsilon^2
\ln [f_1/\epsilon]$, where $\lambda \equiv 10 \sqrt{2} \alpha \beta
\gamma^{3/2}$ and where the function $f_1$ is defined implicitly
(notice that $f_1$ is not defined for extremely small values $\epsilon
< {\rm e}/\lambda \sim 0.03$). This function $f_1 (\epsilon)$ is a
monotonically increasing function of the variable $\epsilon$.
Similarly, the second constraint [\ref{eq:elimit}] says that $N > f_2
(\epsilon) = N_1/\fun(\epsilon)$, where the function $f_2(\epsilon)$
is a monotonically decreasing function of $\epsilon$. Since $N$ must
be greater than both $f_1$ and $f_2$ for all values of $\epsilon$, a
lower bound on $N$ occurs at the crossover point where $f_1 (\epsilon)
= f_2 (\epsilon)$. This lower bound on $N$ is a solution to the
equation 
\be 
N = \lambda \, \bigl[ \fun^{-1} (N_1/N) \bigr]^2 \,  
\ln \Bigl[ {N \over \fun^{-1} (N_1/N)} \Bigr] \,  . 
\ee 
This bound holds for {\it all} values of the star formation
efficiency. To evaluate this bound, we only need to specify the
parameter $\lambda$ (which encapsulates our uncertainties regarding
cluster formation) and the function $\fun$ (which is determined by 
the escape of stars from the cluster during gas removal). For a 
representative case of $\lambda = 80$ ($\alpha=1$; $\beta=2=\gamma$)
and $\fun = (2 \epsilon - \epsilon^2)$, we plot the resulting curves 
$f_1 (\epsilon)$ and $f_2(\epsilon)$ in Figure 1. The intersection 
point determines the constraint $N > N_\star \approx 58$. 

Possible uncertainties in our lower bound $N_\star$ arise from the
form of the function $\fun (\epsilon)$ and the value of $\lambda$.
Fortunately, the result is relatively insensitive to these choices.
If we use alternate fits for the function $\fun$ (Adams 2000), we
obtain essentially the same bound.  For example, the cruder
approximation $\fun$ = $\sqrt{\epsilon}$ changes the crossover point
only by $\Delta N_\star$ $\le$ 1.  Our result also depends only weakly
on the value of $\lambda$; in the limit $N_1 \ll N$, the bound obeys
the scaling law $N_\star \sim \lambda^{1/3}$ (up to a logarithmic
correction), so the constraint is not overly sensitive to $\lambda$.
Figure 1 illustrates this property by plotting alternate curves for
$f_1(\epsilon)$ using $\lambda$ = 40 and 160. The intersections occur
at $N$ = 47 and 73, respectively, which are close to the values
predicted by the $\lambda^{1/3}$ scaling law. Given these
uncertainties and the sharpness of the minimum shown in Figure 1, we
adopt $N_\star$ = 100 as the effective lower limit on the number of
stars required for a system to be a cluster.

This analysis defines a critical value $\ec$ of the star formation
efficiency, i.e., the value corresponding to the crossover point $f_1
(\epsilon)$ = $f_2 (\epsilon)$. For the typical case defined above,
this critical value $\ec \approx 0.4$. For systems with high star
formation efficiency, $\epsilon > \ec$, the constraint of equation
[\ref{eq:nlimit}] dominates and the minimum number of cluster stars is
determined by making the dynamical relaxation time sufficiently
long. For systems with low star formation efficiency, $\epsilon <
\ec$, cluster survival depends on having enough stars remaining
gravitationally bound after gas removal, as enforced by equation
[\ref{eq:elimit}].  Notice also that the constraints shown in Figure 1
define a relatively sharp minimum value of $N$. For values of the star
formation efficiency $\epsilon$ that depart from the critical value
$\ec$, the constraints on $N$ are considerably more restrictive.

The constraints on $N_\star$ derived here depend on both cluster
formation parameters and the function $\fun (\epsilon)$, which, in
turn, depends on the gas structure of the cluster and the distribution
of stellar velocities. We have used basic considerations of cluster
formation and typical parameter values to estimate $N_\star$. However,
all steps of this calculation can be improved. Although the value of
$N_\star$ is thus subject to some uncertainty, the existence of a
limiting value $N_\star$ is not in question -- {\it stellar systems
with too few members will not behave like clusters in a dynamical
sense}.  Such groups lack a collisionless phase and quickly evolve
toward evaporation and dispersal.

The minimum number of stars $N_\star$ $\approx$ 100 is approximately
the same as that of the smallest clusters considered in the analysis
of the BC91 survey. This result also makes sense: Systems with $N \ll
N_\star$ evolve so rapidly and are so dim that they would have little
chance of being included in an observational survey of this type. As a
result, such observational surveys come close to providing a realistic
distribution of the stellar systems that can rightfully be considered
as open clusters (according to this constraint).  We can use the
typical age of open clusters ($\sim$100 Myr; BC91, BT87) to find
another consistency check on this argument: Because a typical cluster
lasts for $\sim$100 relaxation times before significant evaporation
(BT87), the cluster must have a relaxation time of $\sim$1 Myr. This
requirement, in turn, implies a lower bound of $N \ge 160$. These
considerations thus suggest that stellar systems with $N < N_\star
\approx 100$ are not true clusters in that they evolve and disperse
much more quickly than open clusters.  These groups should be
considered as a different type of astronomical system.

\section{EFFECTS OF GROUPS AND CLUSTERS ON STAR FORMATION}

In the previous sections, we made a dynamical distinction between
clusters (large systems with $N>N_\star$) and smaller groups
($N<N_\star$).  We now discuss different ways that groups and clusters
affect the star formation process. In particular, we consider
scattering interactions involving cluster/group members, supernova
explosions, and the background ultraviolet radiation field provided by
the cluster (these results are summarized by Fig. 2).

\subsection{Dynamics and Stellar Scattering Interactions} 

To illustrate the different dynamical effects that groups and clusters
exert upon star formation, we first consider the early evolution
of a stellar system near our estimated boundary at $N$ = $N_\star$ =
100 members. If the cluster forms out of the collapse of a large
molecular cloud core with size $R_0$ = 1 pc and effective sound speed
$a$ = 1 km/s, the formation time is a few million years. With $N$ =
100, the initial value of $Q$ = 7.5 if we assume $\epsilon = 1/2$ so
that the cluster is half gas and half stars. With $v \approx$ 1 km/s
and the typical stellar mass $m_\ast$ $\approx$ 0.5 $M_\odot$, the
total mass is initially 100 $M_\odot$ and the virial size (given by
eq. [\ref{eq:vofr}]) is $R = GM/v^2$ = 0.43 pc (about $R_0$/2).  The
crossing time in this state is 0.43 Myr and the relaxation time is 3.2
Myr. Because the cluster takes 1--2 Myr to form and another 1--2 Myr
to disperse its gas content, it experiences only about one relaxation
time while it remains embedded. After gas removal, the cluster retains
$N \approx$ 75 stars (using $\fun$ = $2 \epsilon - \epsilon^2$) and
the relaxation time drops to 0.25 Myr.  After 25 Myr of additional
evolution, 100 times this starting relaxation time, the group loses
most of its members through evaporation and becomes highly compromised
as a stellar system. It would be impossible to observationally
identify as a cluster. Over its $\sim$25 Myr of evolution, the group
has an average stellar density less than 50 pc$^{-3}$.

Given the above evolutionary picture of our transition-sized cluster,
we consider the possible effects that the system has on its
constituent forming stars and young solar systems. During the
formation stage of the cluster, the typical separation between forming
stars is $\Delta r \approx (4\pi/3N)^{1/3} R \approx$ 0.35 pc. If we
assume that individual stars form through the inside-out collapse of a
centrally condensed structure (as in Shu 1977; Adams, Lada, \& Shu
1987), the region containing a typical stellar mass ($0.5 M_\odot$)
extends over $r_\infty$ = 0.027 pc, $\sim13$ times smaller than the
mean separation (where we assume $a=0.20$ km/s for the individual
infall region). The mean separation between forming stars is much
larger than the size of their protostellar infall regions and
interaction effects are minimal. Similarly, the tidal radius due to
the tidal forces exerted on an individual star forming site by the
gravitational potential of the background cluster is given by $r_T =
\eta (M_\ast/M_{\rm clust})^{1/3} R_0$, where the constant $\eta$
depends on the geometry of the region. For $R_0$ = 1 pc, $M_\ast = 0.5
M_\odot$, $M_{\rm clust}$ = 100 $M_\odot$, and $\eta$ = 1, we thus
obtain $r_T$ $\approx$ 0.17 pc, which is much larger than the size
$r_\infty$ of a protostellar infall region. Tidal effects are thus
small in these transition-sized groups.

An important channel for clusters to affect star formation is through
stellar encounters within the cluster. Such encounters could lead to
binary capture, disk disruption, or changes in planetary orbits. These
effects require young solar systems to experience disruptive close
encounters; keep in mind that all solar systems experience more
distant encounters that lead to dispersal of the cluster.  Let
$\sigma_{200}$ denote the cross section for a close encounter in units
of (200 AU)$^2$ or $\sim9 \times 10^{-7}$ pc$^2$; this is a typical
cross section required for an encounter to force binary capture or to
strongly disrupt a young solar system (e.g., Heller 1993; Ostriker
1994; Laughlin \& Adams 1998, 2000; Kroupa, Petr, \& McCaughrean
1999). For our own solar system, for example, a cross section of (200
AU)$^2$ is the value required to eject Neptune, give Uranus an orbital
eccentricity $e > 0.75$, and/or randomize the orbital inclination
angles of the giant planets (Adams \& Laughlin 2001). In our
transition-sized cluster, the probability $P_D$ for a disruptive close
encounter is given by $P_D \approx \langle n \rangle \sigma_{200} v
(\Delta t)$, where the mean density $\langle n \rangle \approx$ 50
pc$^{-3}$, $v \approx 1$ km/s, and $\Delta t \approx 25$ Myr. The
probability is thus $P_D \approx$ $10^{-3}$ $\sigma_{200}$; the
corresponding odds of a disruptive encounter taking place within the
expected lifetime of the cluster is only $\sim$1 in 1000 (for
$\sigma_{200}$ = 1). Since the cluster contains only 100 stars, the
chances of any disruptive close encounters occurring are about 1 out
of 10. As a rule, more distant encounters disperse the cluster before
disruptive close encounters can greatly alter the constituent solar
systems.

The above considerations suggest that stars forming within
transition-sized clusters ($N=N_\star=100$) experience minimal
dynamical effects from their cluster environment. Interactions between
protostellar infall regions are rare and tidal influences are small.
After solar systems are made, the odds of binary capture or severe
disruption are low, only about 1 part in 1000 per star (1 out of 10
per cluster).  Stars forming within larger clusters ($N \gg N_\star$)
experience the aforementioned dynamical effects with high probability,
whereas stars forming within smaller groups are less likely to
experience such effects.  As a result, our estimated boundary between
groups and clusters (at $N_\star$ $\approx$ 100) also represents an
effective boundary between stellar systems that have a dynamical
impact on forming solar systems (clusters) and smaller systems that 
do not (groups). 

We can illustrate the transition from groups to clusters by deriving a
rough scaling law for $P_D$.  The probability $P_D$ for a close
encounter depends on the mean stellar density $\langle n \rangle$ and
the total lifetime $\Delta t$ of the system.  Using equations
[\ref{eq:trelax} -- \ref{eq:qrelax}], we find the scaling laws
$\langle n \rangle$ $\propto$ $N/R^3$ and $\Delta t$ $\propto$ $(R/v)
N / \ln N$. Putting these results together, we find that the
probability for disruptive encounters takes the form $P_D \propto
N^\mu$, where the index $\mu \approx 2$. According to this relation,
stars living in smaller systems with $N = 30$ are less likely to
experience disruptive encounters by an order of magnitude ($P_D
\approx 10^{-4}$). In these small systems, chances are good (about 1
out 300) that no stars ever experience disruptive encounters.  For
larger systems with $N = 300$, the probability (per star) of
disruptive close encounters increases to $P_D \approx 10^{-2}$
(1\%). For still larger systems with $N = 1000$, the probability of a
disruptive encounter becomes significant, $P_D \approx 0.1$. In these
latter systems, perhaps 100 out of the 1000 cluster members could
experience significant disruption through a close encounter (see
Fig. 2).

This discussion implicitly assumes that the stellar systems under
consideration are large enough (in $N$) so that we can make
statistical arguments for the evolution (in phase space) of the
individual stars. In sufficiently small stellar systems, however, the
dynamics of a given star is dominated by a few close encounters rather
than many distant encounters and hence the scaling laws used here no
longer apply. The criterion for most of the scattering to be due to
weak encounters is $R \gg \bmin = G m_\ast/v^2$, where $R$ is the
cluster size and $m_\ast$ is the typical stellar mass. Using equation
[\ref{eq:vofr}], this requirement reduces to $N \gg 1$, which is met
by most systems. Using standard formalism (BT87), we can make this
requirement more quantitative: The typical velocity perturbation for a
stellar encounter at impact parameter $b$ is given by $\delta v/v$
$\sim$ $\bmin/b$.  Suppose we want to calculate the group size $N$
such that the velocity perturbations are small, specifically $\delta
v/v < \delta_0$, for at least half of the encounters (for a given
choice of $\delta_0$).  Half of the encounters have impact parameters
$b < b_{1/2}$ where $b_{1/2}$ is given by $\ln (R/b_{1/2}) = 0.5 \ln
(R/\bmin)$, i.e., $b_{1/2} = (R \bmin)^{1/2}$. We thus require
$\bmin/b_{1/2}$ $<$ $\delta_0$, which implies $\bmin < \delta_0^2
R$. Using the definition of $\bmin$ and equation [\ref{eq:vofr}], this
constraint can be rewritten in terms of cluster size: $N \delta_0^2 >
1$. For example, if we want at least half of the stellar encounters to
have velocity perturbations $\delta v/v$ $<$ 0.20 = $\delta_0$, we
need a cluster size of at least $N > \delta_0^{-2}$ = 25; similarly,
if we require half the encounters to have $\delta v/v$ $<$ 0.30 =
$\delta_0$, we would need $N > \delta_0^{-2} \approx 11$.  If $N$ is
too small, the evolution of the stellar aggregate is dominated by a
few strong encounters and a statistical description (based on many
weak encounters) breaks down. This result implies an effective lower
boundary for groups; although the boundary is not sharp, systems with
$N < 10$ are dominated by a few hard collisions and hence their
behavior depends sensitively on the initial values of the phase space
variables (see Retterer 1979, BT87; for dynamical simulations of small
($N<10$) systems, see Sterzik \& Durisen 1995, Bonnell et al. 1997).

\subsection{The Probability of Supernovae} 

Another way for a star forming environment to affect its constituent
stars is through supernova explosions. These energetic events can
disrupt star forming regions and remove gas from young clusters;
supernovae have also been invoked as a way to trigger star formation
(e.g., Cameron \& Truran 1977; Boss \& Foster 1998). In the present
context, we argue that groups have little chance of experiencing a
supernova, whereas clusters will often be subjected to their
destructive effects.

To support the above claim, we find the probability $P_{SN}$ that a
stellar system will be subjected to a supernova explosion, as a
function of the number $N$ of stars in the system.  Only stars more
massive than $M_{SN} \approx$ 8 $M_\odot$ explode at the end of their
nuclear burning lives.  The fraction $f_{SN}$ of stars that are
massive enough to explode ($M_\ast > M_{SN}$) depends on the stellar
IMF and is $f_{SN} \approx 0.004$ (see Binney \& Merrifield 1998 for a
discussion of values for $M_{SN}$ and $f_{SN}$).  Next we assume that
the IMF is independent of the size $N$ of the stellar system. To
calculate $P_{SN}$, we imagine picking $N$ stars at random from the
stellar mass distribution. The probability that a given star will {\it
not} be massive enough to explode is ${\widetilde p} = 1 - f_{SN}
\approx 0.996$. The probability that a system of $N$ stars will not
contain an exploding star is thus $({\widetilde p})^N$. Finally, the
probability that a stellar aggregate (with $N$ members) does contain a
progenitor star massive enough to explode is given by 
\be 
P_{SN} (N) = 1 - {\widetilde p}^N = 1 - \bigl [1 - f_{SN} \bigr ]^N \, . 
\label{eq:snprob} 
\ee
Equation [\ref{eq:snprob}] gives the likelihood for supernovae
to occur in a stellar system of size $N$.  For stellar groups, as
defined in \S 3 with $N < N_\star = 100$, the probability of a
supernova is low: $P_{SN} < 0.33$. A natural break-even point between
systems with supernovae and those without occurs where $P_{SN}
(N_{SN})$ = 0.5; the critical number of stars is $N_{SN} \approx 170$.
Larger clusters thus have an appreciable chance of containing stars
large enough to explode as supernovae. Keep in mind that the boundary 
is not perfectly sharp -- stellar aggregates follow a smooth 
probability distribution (given by eq. [\ref{eq:snprob}]; see Fig. 2). 

For supernovae to affect star or planet formation, the stellar system
must live long enough for massive stars to develop iron cores and then
explode; we thus need a cluster lifetime $\Delta t > 10$ Myr. As
discussed in \S 4.1, a transition-sized cluster with $N = N_\star =
100$ is expected to live for $\Delta t \approx 25$ Myr. As a result,
clusters that are large enough (in $N$) to contain massive stars with
high probability are also sufficiently long-lived for their massive
stars to evolve and die while the cluster remains intact.

To summarize, small stellar systems are unlikely to have stars large
enough to explode as supernovae. The boundary between small systems
with no supernovae and larger systems with supernova explosions is
about $N_{SN} \approx 170$. This boundary is roughly coincident with
the boundary between groups and clusters ($N_\star \approx 100$) as
defined by dynamical considerations (\S 3). As a result, stellar
groups will not generally contain supernovae, and will not be
subjected to their disruptive effects nor the possibility of supernova
triggers for star formation.  Larger clusters often contain supernovae
and can experience both their destructive and (possibly) constructive
effects. Throughout this discussion, we assume that the masses of
forming stars obey an IMF that is independent of the system size;
statistics alone then imply that clusters generally contain massive
stars whereas groups generally do not. This distinction becomes even
sharper if massive stars form preferentially within larger clusters as
some authors have conjectured (e.g., Testi, Palla, \& Natta 1999).

\subsection{UV Radiation Fields Provided by Clusters vs Groups} 

External radiation fields from the background environment (the group
or cluster) can have a substantial impact on the star formation
process.  For example, radiation fields can remove gas from
circumstellar disks and thereby suppress both disk accretion and
planet formation (Shu, Johnstone, \& Hollenbach 1993; Hollenbach et
al. 1994; St{\"o}rzer \& Hollenbach 1999). External radiation fields
can also play a role in ending the protostellar infall phase (e.g.,
Hester et al. 1999). These processes are driven mostly by the
ultraviolet (UV) portion of the radiation field, which is dominated by
the most massive stars in the system.  As in the case of supernovae,
the shape of the stellar IMF dictates that massive stars are rare
except in sufficiently large systems. As a result, solar systems
forming within large stellar aggregates (clusters) receive an
appreciable contribution of UV radiation from their background
cluster; solar systems forming within small stellar groups receive
little UV radiation from the background. 

To substantiate this claim, we estimate the UV radiation field
provided by a stellar aggregate, as a function of the number $N$ of
stars in the system. We follow a previous calculation of the UV field
for the expected conditions experienced by our own solar system during
its planet formation epoch (Adams \& Laughlin 2001). This calculation
finds the expectation value for the ionizing ultraviolet flux from a
background stellar system, i.e., 
\be
\fuv = 1.6 \times 10^{12} {\rm cm}^{-2} {\rm sec}^{-1} \, 
\Bigl( {N \over 2000} \Bigr) 
\Bigl( {R \over 1 {\rm pc} } \Bigr)^{-2} \, , 
\label{eq:uvcluster} 
\ee
where $N$ is the number of stars and $R$ is the cluster size. This
expression was obtained by making two averages: We first integrate
over a typical stellar orbit through the cluster to find the mean flux
impinging upon a given solar system due to the massive stars, which
provide the UV flux and are assumed to reside at the center. We also
integrate over the stellar IMF, weighted by the UV luminosity as a
function of stellar mass, to find the ionizing UV flux as a function
of the number $N$ of stars in the system (Adams \& Laughlin 2001).
For comparison, the ionizing UV luminosity of a 1 $M_\odot$ star
during its pre-main-sequence phase cannot be larger than about $L_{uv}
\approx 10^{41}$ sec$^{-1}$ (Gahm et al. 1979). The corresponding UV
flux from the star is given by 
\be 
F_{uv \ast} = 3.5 \times 10^{13} {\rm cm}^{-2} {\rm sec}^{-1} \, 
\Bigl( {r_\ast \over 1 {\rm AU} } \Bigr)^{-2} \, ,
\label{eq:uvstar}
\ee
where $r_\ast$ is the radial coordinate centered on the star. 

One measure of the importance of the local UV radiation background is
the total number of photons intercepted by circumstellar disks. These
disks actively form planets during their first $\sim10$ Myr (Lissauer
1993). Ultraviolet radiation acts to drive a wind from the disk
surface, remove gas from the disk, and eventually compromise planet
formation.  For a typical disk, we want to compare the number of
ionizing UV photons provided by its background cluster
(eq. [\ref{eq:uvcluster}]) with the UV radiation intercepted from its
central star (eq. [\ref{eq:uvstar}]). The disk is embedded in the
background UV radiation field and both sides are exposed; the disk
thus receives UV photons from the cluster at a rate 
\be 
\Phi_{uv} = 2 \pi R_{\rm d}^2 \, \fuv \approx 10^{41}
{\rm sec}^{-1} \Bigl( {N \over 100} \Bigr) 
\Bigl( {R \over 1 {\rm pc} } \Bigr)^{-2} \, , 
\label{eq:uvtotal} 
\ee
where $R_{\rm d}$ $\approx$ 30 AU is the radial size of the disk. The
nominal value of $\Phi_{uv}$ (which has been rescaled to $N=100$) is
equal to the total production rate of ionizing UV photons from a 1
$M_\odot$ star ($\sim 10^{41}$ sec$^{-1}$). Sufficiently large
clusters (those with $N > 100$) produce enough ionizing UV radiation
to dominate the UV flux intercepted by circumstellar disks; smaller
groups ($N<100)$ have smaller UV backgrounds and circumstellar disks
are primarily irradiated by their central stars (see Fig. 2).

We now derive the break-even point between stellar systems that are
large enough to dominate the ionizing UV field experienced by a
circumstellar disk and smaller systems in which individual stars
provide most of the UV to their disks. In general, disks intercept
only a fraction of the UV photons generated by their central stars.
Taking into account both disk flaring and scattered (diffuse) photons,
Shu et al. (1993) estimate that nearly 50$\%$ of the UV photons are
intercepted by the disk. The rate of intercepted (intrinsic) stellar
UV photons is thus $\Phi_\ast \approx 5 \times 10^{40}$ sec$^{-1}$ for
a 1 $M_\odot$ star.  The scaling law [\ref{eq:uvtotal}] implies that
stellar systems with $N > 50$ provide more ionizing UV photons to a
circumstellar disk than its central star. According to this criterion
of ionizing UV radiation, the boundary between large stellar systems
(clusters) that affect circumstellar disks and smaller systems
(groups) that do not is $N \approx 50$.

This result was derived using an expectation value for the mean UV
flux from a stellar aggregate. With cluster/group sizes as small as
$N=50-100$, the IMF is not completely sampled by any given stellar
system and individual groups/clusters will experience sizable
fluctuations about the expectation value (eq. [\ref{eq:uvtotal}]).  
In addition, equation [\ref{eq:uvcluster}] was derived without taking
into account attenuation of the UV flux by gas and dust in the
cluster. Since the gas removal time is relatively short (a few Myr)
compared to the expected disk lifetimes and the planet formation times
($\sim10$ Myr), an attenuation correction would not appreciably
change the estimated boundary at $N=50-100$.  Finally, we have only
considered the ionizing radiation in this discussion. Photodissociation 
can be as important as photoionization and should be studied in a more
complete treatment (see Diaz-Miller, Franco, \& Shore 1998;
St{\"o}rzer \& Hollenbach 1999).

\section{CONCLUSION} 

\subsection{Summary of Results} 

In this paper, we have obtained three principal results: 

{\bf [1]} Using the estimated cluster formation rate, we find that
less than 10$\%$ of stars are formed in systems that become open
clusters (\S 2). In this context, only relatively large systems (with
a few hundred stars or more) are considered to be bona fide clusters. 
Of the nearly 90\% of star formation that takes place in other 
environments, perhaps the majority takes place in smaller systems 
-- denoted here as groups -- with $N = 10 - 100$ stars.

{\bf [2]} Viable clusters -- those which survive to be observable as
open clusters -- must have a minimum number $N_\star$ of stars. We
estimate this limiting value by requiring that the cluster formation
time is shorter than the dynamical relaxation time (\S 3) and thereby
find the minimum number of cluster stars: $N_{\star} \approx 100$. 
This critical value $N_\star$ provides an effective boundary between 
clusters ($N > N_\star$) and smaller stellar groups ($N < N_\star$).

{\bf [3]} Small stellar groups and larger clusters affect star
formation in different ways (\S 4). The small fraction ($\sim$1/10) of
stars that form within cluster systems are likely to be affected by
disruptive close encounters (\S 4.1), such as stellar mergers, binary
capture via disks, disruption of planetary orbits, strong interactions
between protostars, and competitive accretion.  Stars forming within
clusters are also influenced by nearby massive stars, through both
supernova explosions (\S 4.2) and background UV radiation fields (\S
4.3). For the majority of stars ($\sim$9/10), however, none of the
aforementioned effects operate with high probability and the
background environment is sufficiently diffuse to allow individual
stars to form in relative isolation.

\subsection{Modes of Star Formation} 

The results of this paper suggest that we must consider star formation
to take place in more modes than has been historically recognized. In
particular, we need to move beyond the traditional dichotomy between
isolated single stars and clusters. The following classes provide a
starting point:

[{\bf I}] Isolated single stars and multiple systems, including
binaries, triples, and other few body systems.  For this class of
systems, the number $N$ of stellar members is less than about 10,
although more typically $N$ = 2 or 3. We could subdivide this class
further into multiple systems and single stars (thus defining a Class
[{\bf 0}]). Although single stars may well be in the minority, this 
zeroeth class would include our solar system. 

[{\bf II}] Groups, consisting of intermediate numbers of stars with
$10 < N < N_\star \sim 100$. The upper limit $N_\star$ is determined
by the criterion that the dynamical relaxation time must be longer
than the formation time of the cluster; the value $N_\star$ $\approx$
100 is thus approximate and depends on manner in which clusters
form. The lower limit at $N \approx 10$ is also approximate and marks
the boundary between few-body systems and larger groups in which
stellar dynamics can be described statistically. This mode of star
formation may be dominant in that most stars may form in such
groups. This work suggests that stars forming within small groups 
are largely decoupled from their immediate environment. 

[{\bf III}] Clusters, consisting of large numbers of stars with $N >
N_\star \sim 100$. This regime corresponds to robust clusters that
live for long times, i.e., systems that can be observationally
identified as open clusters. Stars forming within these systems are
subjected to dynamical effects such as core mergers, binary capture,
and planetary scattering. Clusters have large enough stellar
membership to contain massive stars with high probability; stars
forming in clusters are thus exposed to intense UV radiation and
supernova blast waves.

These three modes of star formation represent the types of stellar
aggregates that a forming star might find itself within. These modes,
or classes, differ from each other in two important respects: [1] They
represent different kinds of stellar systems that exhibit different
dynamical behavior and evolution (cf. \S 2 and \S 3).  [2] They
represent different types of star formation environments that affect
the star formation process in different ways (\S 4). In particular,
cluster environments subject their stellar constituents to a host of
disruptive (and constructive) influences, whereas smaller groups have
relatively little effect on the star formation process.

\subsection{Discussion} 

Perhaps the most important result of this paper is to make a clearer
distinction between groups and clusters. We have made this distinction
in five different ways and thereby obtain five estimates for the
boundary $N_\star$ between groups ($N<N_\star$) and clusters
($N>N_\star$): [1] Clusters are big and bright enough, and live long
enough, to be included in observational surveys (\S 2; $N_\star \sim
300$). [2] Clusters have dynamical relaxation times that are longer
than their formation times (\S 3; $N_\star \approx 60 - 100$). [3]
Clusters are sufficiently dense and long-lived so that disruptive
scattering encounters can affect circumstellar disks and their
planetary progeny (\S 4.1; $N_\star \sim 100$). [4] Clusters have
enough stellar members and live long enough so that supernovae can
affect forming stars (\S 4.2; $N_\star \approx 170$). [5] Clusters
have enough massive stars so that the ionizing UV radiation field
impinging upon forming planetary systems is dominated by the
background cluster rather than the central star (\S 4.3; $N_\star
\approx 50 - 100$). These five determinations are roughly coincident
and imply $N_\star \sim 100$. However, the boundary between groups and
clusters is not perfectly sharp: stellar systems exhibit a continuous
distribution of properties as a function of stellar membership $N$
(see Fig. 2).

These criteria for distinguishing groups and clusters depend on many
different (but sometimes coupled) physical processes: the longevity of
stellar systems, the stellar IMF, the UV mass-luminosity relationship
for stars, the minimum progenitor mass for a supernova, the brightness
of stellar populations, the formation time for clusters, and the
scattering cross sections for solar system disruption. The complicated
interplay between stellar dynamics and stellar physics thus leads to a
relatively clean distinction between groups and clusters, both as
stellar systems (they behave differently as astronomical entities) and
as star formation environments (they affect star formation differently). 

Some ambiguity remains in the relative portions of stars that form in
the three classes defined here. The observational cluster surveys
(e.g., BC91) show that only 10\% of star formation takes place within
{\it the clusters included in the sample}. Although BC91 use a mass
function for clusters that extends down to $m_1$, where $40 M_\odot
\le m_1 \le 120 M_\odot$ for their various models, the accounting for
the lowest mass clusters is not necessarily complete. The
observational surveys could thus be missing small clusters in the
range 100 $\le N \le$ 300. One common explanation for why stars are
not seen in open clusters is that they form within cluster-sized
units, but the units disperse quickly after gas removal. If this
scenario is true, then the cluster-sized units must be small enough to
avoid detection in the surveys (which include young clusters with
$t\sim$3 Myr and all bright clusters). It thus remains possible for a
substantial fraction of star formation to take place in systems with
$N$ in the range 100 $\le N \le$ 300. However, such small stellar
systems disperse rapidly and are thus more like groups than clusters;
in addition, these small systems have relatively little effect on star
formation (see \S 4 and Fig. 2).

The relative amount of star formation that takes place in groups
versus the isolated mode must also be better specified. Using the
current data base of dense cores mapped in ammonia (Jijina, Myers, \&
Adams 1999), we find that most of the dense gas is contained in cores
associated with stellar aggregates with $N > 30$ (safely in the regime
of groups as defined here). Although the data base clearly shows that
more star formation takes place in groups than in the isolated mode,
the heterogeneous nature of the data makes further quantitative
determinations difficult. This issue thus requires further work.

When star formation takes place within groups, the most common result
is complete dispersal of the system on a relatively short time scale,
due to gas removal (a few Myr) and dynamical evolution (10--20 Myr).
For example, Orion A (L1641) contains both groups and more widely
distributed stars. It has been suggested that the distributed
population in L1641 may have been born in larger aggregates like those
seen now, but the groups have already dispersed (Strom et al. 1994).
On the other hand, groups forming in close proximity could merge and
thereby build larger clusters after their initial formative stage. The
formation and evolution of intermediate sized stellar systems ($N \sim
100$) thus constitutes an important area for future work.

Another goal for the future is to construct the distribution $P(N)$
which gives the probability that a star will form within a system of
$N$ members. This work indicates that the probability distribution
contains three principle components: isolated and few-body systems
with $N < 10$, small groups with $10 < N < N_\star$ $\sim100$ (perhaps
containing the majority of forming stars), and a tail of larger
stellar systems representing the clusters with $N > N_\star$
(comprising about 10\% of the population). Although the available data
are not adequate to define a reliable distribution at this time, the
construction of $P(N)$ should become feasible in the near future. 

\acknowledgments

We would like to thank Gus Evrard, Suzanne Hawley, Charlie Lada, Greg
Laughlin, and Frank Shu for useful discussions. We also thank an
anonymous referee and Scientific Editor Steve Shore for many helpful
suggestions that greatly improved the paper. This work was supported
by funding from The University of Michigan and by bridging support
from NASA Grant No. 5-2869.

\clearpage

\figcaption[]{
Two simultaneous constraints required for stellar systems to be large
enough to behave like clusters in a dynamical sense. The constraints
are plotted as a function of star formation efficiency $\epsilon$. The
dashed curve $f_1$ represents the minimum number of stars required for
the dynamical relaxation time of the system to be longer than the
cluster formation time (for the standard choice of parameters leading
to $\lambda$ = 80).  The solid curve $f_2$ represents the minimum
number of stars required for the system to remain bound after gas
removal, where the fraction of stars remaining is given by the
analytic fitting function $\fun$ = $2 \epsilon - \epsilon^2$. Since
both constraints must be satisfied, the minimum number of cluster
stars is determined by the intersection point at $N \approx 60$ (and
$\epsilon \approx 0.38$).  The two dotted curves represent alternate
parameters choices ($\lambda$ = 40 and 160) for the function
$f_1$. Notice that the value of $N$ at the intersection point is not
very sensitive to the choice of $\lambda$: For $\lambda$ = 40 (160),
the intersection occurs at $N$ = 47 (73).  
\label{fig1}}

\figcaption[]{
The transition from groups (small $N$) to clusters (large $N$).  
The curve labeled UV shows the relative fraction of the ionizing 
ultraviolet radiation that is provided by the background environment,
${\cal F}_{uv} \equiv \Phi_{uv} (N)/ [\Phi_\ast + \Phi_{uv}]$ 
(\S 4.3), as a function of the size $N$ of the stellar aggregate 
(horizontal axis).  The curve labeled SN shows the probability
$P_{SN}$ that a stellar aggregate of size $N$ will contain a supernova
(\S 4.2, eq.  [\ref{eq:snprob}]).  The curve labeled $D_1$ shows that
probability that a stellar aggregate will produce at least one
disruptive scattering encounter (\S 4.1); the dashed curve labeled
$D_N$ shows the probability that any given solar system in the cluster
will experience a disruptive scattering encounter. (For comparison,
observational surveys of open clusters include stellar aggregates with
sizes $N$ down to about $N \sim 300$; the minimum number of stars 
necessary for a stellar system to have its dynamical relaxation time 
longer than its formation time is $N_\star \approx$ 100.) 
\label{fig2}} 

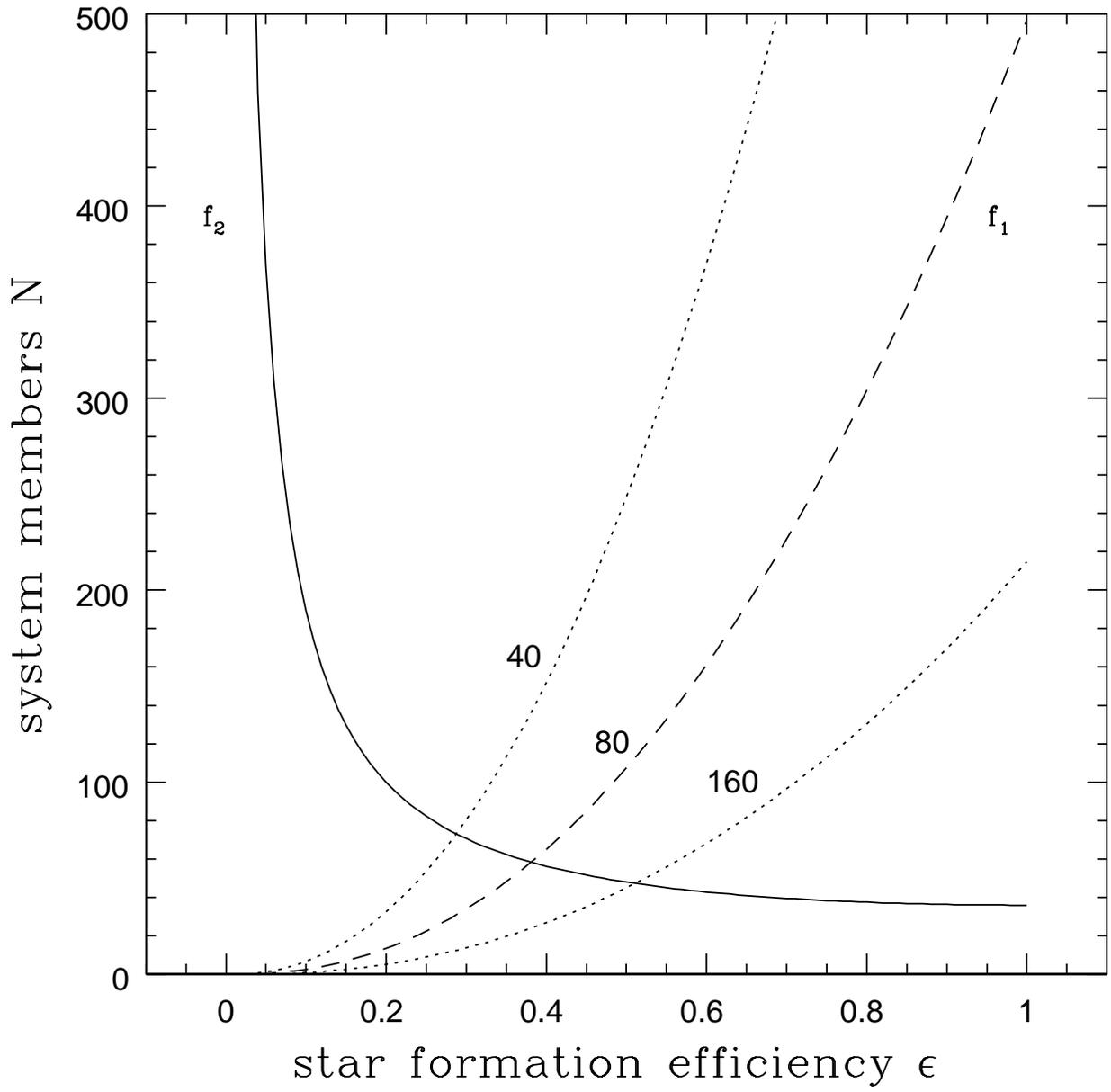
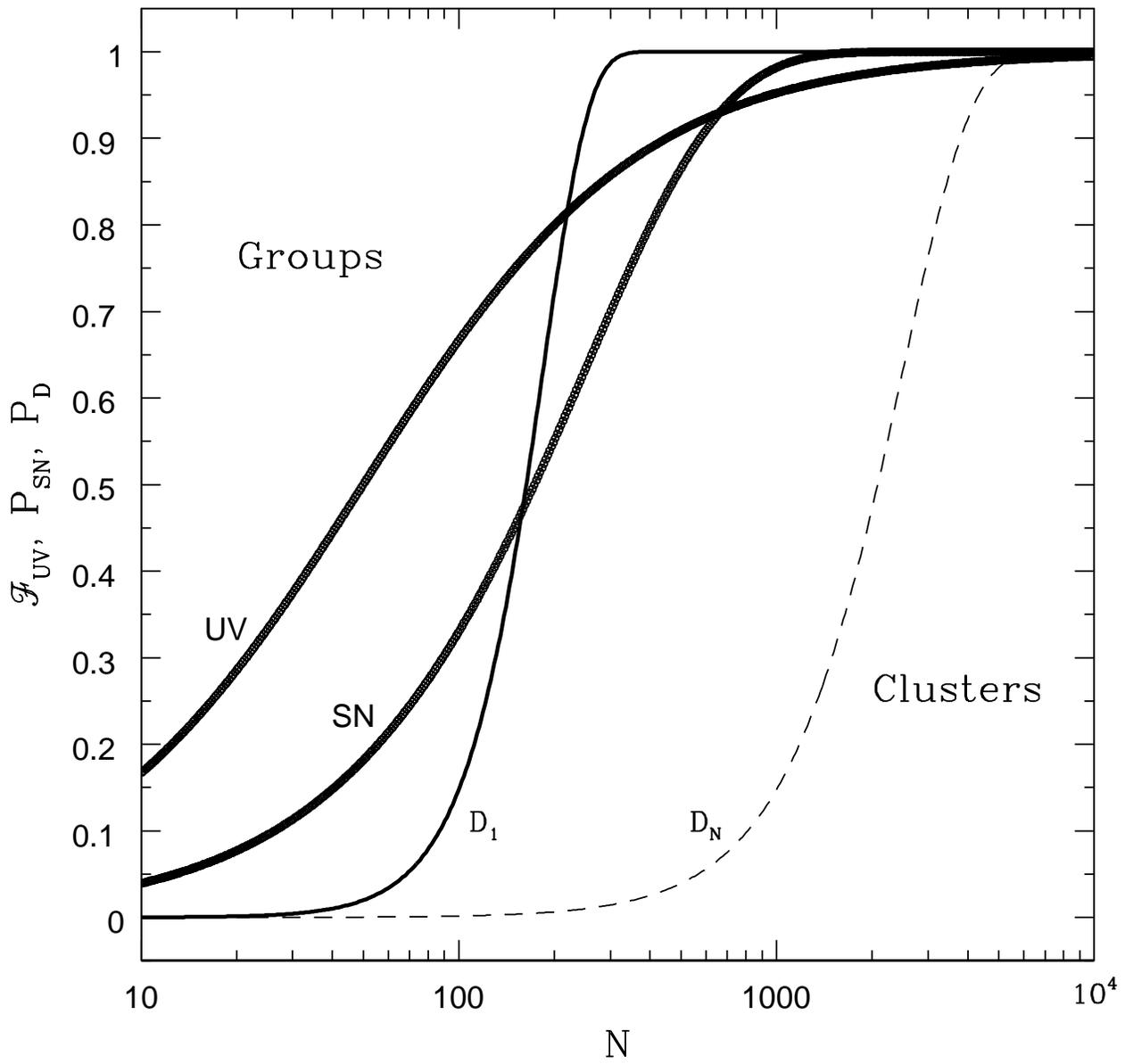
\end{document}